\documentclass[english,twocolumn,aps,prl,showpacs,amsmath]{revtex4}
\usepackage[OT1]{fontenc}
\usepackage[latin1] {inputenc}
\usepackage{graphicx}
\usepackage{amssymb}

\usepackage{bm}
\usepackage{latexsym}
\usepackage{hyperref}
\usepackage{babel}
\makeatother
\begin{document}
\title{Adiabatic solution of Dirac equation of ``graphinos'' in an intense electromagnetic field and emission of high order harmonics 
near the Dirac points}
\author{F.H.M. Faisal}
\affiliation{Fakult\"at f\"ur Physik, Universit\"at Bielefeld, Postfach 100131,
D-33501 Bielefeld, Germany}

\begin{abstract}
We obtain a class of adiabatic solutions of Dirac equation for the charged massless relativistic quasi-particles that arise from the low-energy excitations \cite{foot-1} 
in a 2D graphene sheet,
interacting with an electromagnetic field. The analytic solutions obtained are useful for {\it non-perturbative} investigation of processes in intense laser fields.
As a first example we employ them to predict copious emissions of high order harmonics by THz lasers 
interacting with the occupied states of graphinos in the vicinity of the degenerate Dirac points. The relative intensity of the emitted harmonics is seen to decrease by only about two orders of magnitude  from the 3rd to the 81st  harmonic order, and is characterized by two phenomena of ``revival'' and ``plateau'' formation in the middle and the far end of the spectrum calculated. A preliminary comparison is made for harmonic emission from 2D graphene that reveals a qualitatively different spectrum in the latter case showing a sharp cutoff at an order $n_{cutoff} = \frac{\omega _B}{\omega}$ where $\omega_B=eF_0b/\hbar$ is the so-called Bloch frequency.
\end{abstract}
\maketitle

\section{Introduction}
Graphene is a two-dimensional single layer of carbon atoms in a hexagonal honeycomb arrangement (Fig. 1 (a)). Since its recent controlled synthesis in the laboratory graphene has emerged as an object of much current research interest \cite{rev1, rev2} 
with potentials for wide range of applications in various quantum technologies. A remarkable property of a $2D$ graphene membrane is that its low energy excitations -- near the so-called Dirac points (${K},{K '}$) at the corners of the graphene
Brillouin zone (cf. Fig. 1 (b)) --
are relativistic fermions that are charged like the electrons but massless like the neutrinos. 
They move with a constant velocity $v_{F}$ that makes them fully relativistic at a much lower velocity than the velocity of light $c$ $(v_{F}\approx c/300)$. Also, their anti-particle counterparts are oppositely charged massless holes with opposite helicity \cite{foot-1}. 

The stationary eigenvalues and the spinor eigenfunctions of the Dirac equation of free graphinos as well as 
the wavefunctions in the presence of static electric and magnetic fields, and in crossed electric plus magnetic fields have been obtained recently (see e.g. review\cite{rev2} and \cite{GraCroFie1, GraCroFie2}). They have been used to investigte some of the most remarkable properties of graphinos, such as chiral scattering, confinement, zitterbewegung, and anomalous integer quantum Hall effect -- among others -- that have been observed \cite{rev1, rev2}.
In this paper we consider the adiabatic solutions of the {\it time-dependent} Dirac equation of graphinos interacting with an electromagnetic field of a given pulse shape, peak field strength, central frequency, and polarization. As a first application of the solution obtained, we derive the Dirac expectation value of the graphino current in the presence of the field 
and show that copious emission of high harmonic radiations can occur from the initially occupied states of graphino pseudo-particles near the degenarate Dirac points, by an intense THz laser.
\begin{figure}
\begin{centering}
\includegraphics[scale=0.33]{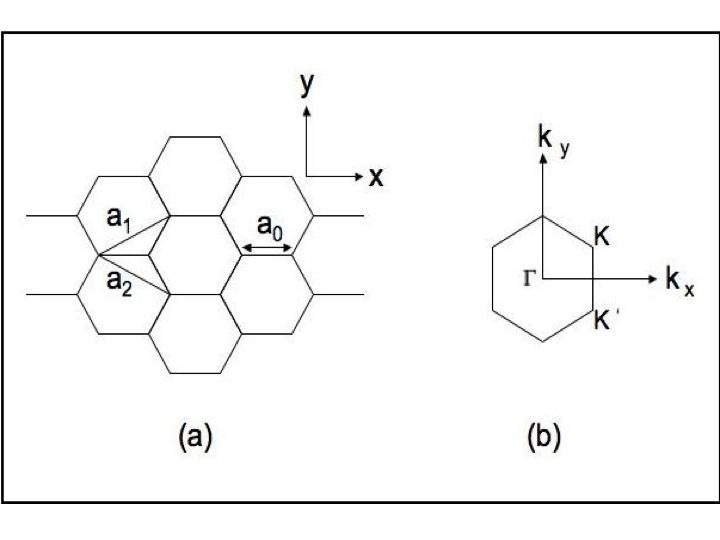}
\par\end{centering}
\caption{\label{Fig. 1}  Structure of graphene: (a) 2D hexagonal honeycomb lattice, $\bm{a}_{1}= a \hat{x} +b \hat{y}, \bm{a}_{2}= a \hat{x} - b\hat{y},
a=\frac{3}{2} a_{0}, b=\frac{\sqrt{3}}{2}a_{0}$; (b) Brillouin zone showing the Dirac points $K(\frac{2\pi}{3 a_{0}}, \frac{2\pi}{3\sqrt{3} a_{0}})$ and $K '(\frac{2\pi}{3 a_{0}}, -\frac{2\pi}{3\sqrt{3} a_{0}})$.}
\end{figure}

\section{Dirac Equation of a Graphino Interacting with a Transverse Electromagnetic Field}
The $2D$ Dirac Hamiltonian of a graphino is given by 
\begin{equation}\label{GraDirHam}
H= v_{F}{\bf \sigma}\cdot {\bf p}_{op}
\end{equation}
where
${\bf \sigma} =(\sigma_{x},\sigma_{y})$ are $2\times 2$ Pauli matrices
\begin{equation}\label{SpiMat}
 \sigma_x = \begin{pmatrix} 0&1\\ 1&0 \end{pmatrix}, 
 \sigma_y = \begin{pmatrix} 0&-i\\ i&0 \end{pmatrix},
\end{equation}
$v_{F}$ is the graphino velocity, and ${\bf p}_{op}\equiv-i\hbar\frac{\partial}{\partial\bf{r}}$, ${\bf r}= (x,y)$.
We consider a general plane wave electromagnetic field, assumed to be elliptically polarized in the graphene plane, to be 
represented by the vector potential
\begin{eqnarray}\label{VecPot}
{\bf A}(t) &=&  A_{0}(\eta)\left({\hat \epsilon}_{x}\cos{(\xi/2)}\cos{\eta} -{\hat \epsilon}_{y}\sin{(\xi/2)}\sin{\eta}\right), \nonumber\\  
\eta &=& (\omega t -\kappa z)
\end{eqnarray}
where $A_{0}(\eta)$ is the pulse envelope,
 $\xi = [0,\pm\pi/2]$ is the ellipticity parameter 
with left  ($+$) or right $(-)$ handedness, 
and ${\hat \epsilon}_{x}, {\hat \epsilon}_{y}$ are unit polarization vectors.
The minimal-coupling prescription yields the Dirac equation of the ``graphino + electromagnetic field'': 
\begin{equation}\label{DynDirEqu}
i\hbar \frac{\partial}{\partial t} \Psi(t) = \left( v_{F}{\bf \sigma}\cdot ( {\bf p}_{op}-\frac{e}{c}{\bf A}(\eta)) \right) \Psi(t)
\end{equation}
\subsection{The unperturbed graphino} 
It will be useful first to briefly consider the eigen-solutions \cite{rev2} of the field-free graphino Dirac equation governed by the Hamiltonian (\ref{GraDirHam}),
\begin{equation}\label{GraDirEqu} 
E\psi_{\bf p}({\bf r}) = v_{F}{\bf \sigma}\cdot{\bf p}\psi_{\bf p}({\bf r}) 
\end{equation}
where $E$ is the energy and $\bm{p} =\hbar \bm{q}$ is the quasi-momentum measured from the Dirac point $\bm{K}$ (or $\bm{K '}$). One 
substitutes
$\psi_{\bf p}({\bf r}) = e^{\frac{i}{\hbar}{\bf p}\cdot{\bf r}} u_{\bf p}$, where
\begin{equation}
u_{\bf p} = \begin{pmatrix}  a\\ b \end{pmatrix}
\end{equation} 
is a two component spinor, in Eq. (\ref{GraDirEqu}) to get
\begin{eqnarray}\label{GraEneEqu}
E u_{\bf p} &=& v_{F} {\bf \sigma}\cdot {\bf p} u_{\bf p}\nonumber\\
&=& v_{F}p \begin{pmatrix} 0&e^{-i\phi_{p}}\\ e^{i\phi_{p}}&0\end{pmatrix}u_{p}
\end{eqnarray}
where $\phi_{p}$ is the polar angle of the 2D momentum ${\bf p}$.
This gives a $2\times 2$ matrix equation for the components $a$ and $b$ of the spinor $u_{\bf p}$
\begin{equation}
E \begin{pmatrix} a\\b\end{pmatrix} = v_{F}p \begin{pmatrix} 0&e^{-i\phi_{p}}\\ e^{i\phi_{p}}&0\end{pmatrix}\begin{pmatrix}a\\b\end{pmatrix}
\end{equation}
This can be readily solved
to obtain the two eigenvalues
\begin{equation}\label{GraEigVal}
E^{(\pm)} = \pm v_{F} p 
\end{equation}
and the corresponding normalized eigenspinors,
\begin{equation}\label{GraEigVec}
u_{p}^{(\pm)}= \frac{1}{\sqrt{2}}\begin{pmatrix} e^{-i\phi_{p}/2}\\ \pm e^{i\phi_{p}/2}\end{pmatrix}.
\end{equation}
Thus the normalized eigenstates of Eq. (\ref{GraDirEqu}) are 
\begin{eqnarray}\label{GraEigSol}
\psi_{\bf p}^{(s)}({\bf r}) &=& \frac{1}{\sqrt{{\it A}}} e^{\frac{i}{\hbar}{\bf p}\cdot {\bf r}} u_{\bf p}^{(s)}
\end{eqnarray} 
with eigenvalues $E^{(s)} (\bm{p})= s v_{F} p$,  where ${\it A}$ is the quantization area, and $s=\pm 1$ are the two components of the graphino psedo-spin, corresponding to the limiting case of the upper and the lower band of graphene, respectively. 

\subsection {Adiabatic solutions of graphino Dirac equation in a laser field}
We may now proceed to obtain a class of useful solutions of the time-dependent Dirac equation (\ref{DynDirEqu}) for the interacting system ``graphino + laser field''. We first  make the following adiabatic ansatz for the wavefunction (e.g. \cite{schiff}) 
\begin{equation}\label{AdiAns}
\Psi_{\bf p}({\bf r},t) = \frac{1}{\sqrt{\it A}} e^{-\frac{i}{\hbar}\int^{t} E(t') dt'}  e^{\frac{i}{\hbar}{\bf p}\cdot{\bf r}}  u_{p}(t)
\end{equation}
where $E(t)$ and $u_{\bf p}(t)$ are time dependent unknowns to be determined.
To this end
we substitute (\ref{AdiAns}) in Eq. (\ref{DynDirEqu}) and obtain
\begin{eqnarray}
&&E(t) u_{p}(t) +   i\hbar \frac{\partial}{\partial t} u_{p}(t) \nonumber\\
&=& \left( v_{F}{\bf \sigma}\cdot ( {\bf p} -\frac{e}{c}{\bf A}(\eta)) \right) u_{p}(t)
\end{eqnarray}
Next, projection from the left with the hermitian adjoint spinor $u_{p}^{\dagger}(t)$ gives 
\begin{eqnarray}\label{AdiEqu}
&&E(t) u_{p}^{\dagger}(t)u_{p}(t) +   i\hbar u_{p}^{\dagger}(t) {\dot u}_{p}(t)\nonumber\\
&=& u_{p}^{\dagger}(t)\left( v_{F} {\bf \sigma}\cdot ( {\bf p} -\frac{e}{c}{\bf A}(\eta)) \right) u_{p}(t)
\end{eqnarray} 
where we have denoted ${\dot u}_{p}(t) = \frac{\partial}{\partial t}u_{p}(t)$.
Comparing the last term in (\ref{AdiEqu}) with the unperturbed equations 
(\ref{GraEneEqu}) and (\ref{GraEigVec}) 
we make the ansatz
\begin{equation}\label{TimDepSpi}
u_{\bf p}(t) \equiv u^{(s)}_{{\bf p}(t)} =\frac{1}{\sqrt{2}}\begin{pmatrix} e^{-i\phi_{p(t)}/2}\\ s e^{i\phi_{p(t)}/2}\end{pmatrix},  
\end{equation}
where we have defined 
${\bf p}(t) = ({\bf p} -\frac{e}{c}{\bf A}(\eta)), 
p(t) = |{\bf p}(t)|$,
and $\tan\phi_{p(t)} = p_{y}(t)/p_{x}(t)$.
It is easily seen by direct calculation that,
\begin{equation}\label{OrtNor}
u^{(s)\dagger}_{p(t)}u^{(s')}_{p(t)} = \delta_{s,s'}
\end{equation} 
By calculations similar to that shown above for the unperturbed case but with the time-dependent spinors defined by Eq. (\ref{TimDepSpi}) we also find  
\begin{eqnarray}\label{EquAdi}
\left( v_{F} {\bf \sigma}\cdot ( {\bf p} -\frac{e}{c}{\bf A}(\eta)) \right) u_{p}^{(s)}(t)
&=& s v_{F}|{\bf p}(t)| u_{p}^{(s)}(t)
\end{eqnarray}
Moreover, since 
\begin{equation}\label{Ide1}
{\dot {u}}_{p}^{(s)}(t)=\frac{1}{\sqrt{2}}\begin{pmatrix} e^{-i\phi_{p(t)}/2}(-\frac{i}{2}{\dot{\phi}}_{p(t)})\\ s e^{i\phi_{p(t)}/2}(\frac{i}{2}{\dot{\phi}}_{p(t)})\end{pmatrix},  
\end{equation}
and
\begin{equation}\label{Ide2}
u_{p}^{(s)\dagger}(t) =\frac{1}{\sqrt{2} }( e^{i\phi_{p(t)}/2}, s e^{-i\phi_{p(t)}/2}),  
\end{equation}
therefore the inner product 
\begin{eqnarray}\label{InnPro}
u_{p}^{(s)\dagger}(t) {\dot u}_{p}^{(s)}(t)=\frac{1}{4}( -i{\dot{\phi}}_{p}(t) +i{\dot{\phi}}_{p}(t))&=&0
\end{eqnarray}
Thus, substituting Eqs. (\ref{OrtNor}), (\ref{EquAdi}) and (\ref{InnPro}) in Eq. (\ref{AdiEqu})
it is seen that Eq. (\ref{AdiEqu}) is fully satisfied provided 
\begin{equation}\label{AdiEigVal}
E(t)= s v_{F}|{\bf p}(t)|
\end{equation}
Hence, finally, we obtain the desired adiabatic solution of the time-dependent Dirac equation (\ref{DynDirEqu})
\begin{eqnarray}\label{AdiSol}  
\Psi^{(s)}_{{\bm p}}({\bf r},t)&=&\frac{1}{\sqrt{A}} 
e^{-\frac{i}{\hbar} s v_{F} \int^{t} |{\bf p}(t')| dt'} e^{\frac{i}{\hbar}{\bf p}\cdot {\bf r}}
u^{(s)}_{{\bf p}(t)}
\end{eqnarray}
with $s=\pm 1$.
They fulfil the biorthonormal conditions
\begin{eqnarray}\label{ComRel}
\int \Psi^{(s')\dagger}_{\bf p'}({\bf r},t)\Psi^{(s)}_{\bf p}({\bf r}, t) d^{2}r &=&\delta({\bf p} -{\bf p'})\delta_{s,s'}\nonumber\\ 
\sum_{\bf p}\sum_{s=\pm} \Psi^{(s)}_{\bf{p}}({\bf r'},t) \Psi^{(s)\dagger}_{\bf p}({\bf r},t) & =& \delta({\bf r} -{\bf r'})
\end{eqnarray}
and thus form a complete set of functions that can be used to represent an adiabatic state $\left|\Phi(t)\right\rangle$ 
satisfying any given initial condition $\left|\Phi(t_{0})\right\rangle$ at a time $t_{0}$, in the form
\begin{equation}
\left|\Phi(t)\right\rangle = \sum_{\bf {p}}\sum_{s=\pm} \langle\Psi^{(s)}_{\bf{p}}(t_{0})|\Phi(t_{0})\rangle \left|\Psi^{(s)}_{\bf{p}}(t)\right \rangle
\end{equation}

\section{High Harmonic Emission from Graphino}
As a first application of the solution obtained above, we  
consider generation of higher order harmonics from graphinos near the Dirac points by intense laser fields. 
Emission of high harmonic of radiation might be expected in analogy with the earlier theoretical investigations within the elaborate Floquet-Bloch theory
for crystal surfaces and graphene \cite{fai-kam, fai-kam-saszuk, moiseyev}) or from calculations with two-state models of semiconductors
\cite{golde}, as well as from experimental observations of higher order harmonic emissions from surfaces and bulks \cite{vonderlinde, norreys, dimauro} by high intensity infrared laser fields. In view of the zero-mass of graphinos we may expect their strong effective coupling with low frequency laser fields and consequent copious emission of high harmonics, even by moderately intense lasers. To show this theoretically we next obtain the quantum expectation value of associated current operator in an electromagnetic field and calculate the relative harmonic intensities from a state occupied in the vicinity of the Dirac points, for the case of a THz laser.
\subsection{Graphino current driven by an electromagnetic field}
From  Eq.(\ref{GraDirEqu}) we have
\begin{eqnarray}\label{Dir}
\frac{\partial}{\partial t} \Psi_{\bf p}(t) &=& -v_{F} {\bf \sigma}\cdot\nabla\Psi_{\bf p}(t)\nonumber\\
&=& -v_{F} {\bf \sigma}\cdot\nabla\Psi_{\bf p}(t)
\end{eqnarray}
and from its hermitian conjugate
\begin{eqnarray}\label{ConDir}
\frac{\partial}{\partial t} \Psi^{\dagger}_{\bf p}(t) &=& -v_{F} {\bf \sigma}\cdot\nabla\Psi^{*}_{\bf p}(t)\nonumber\\
&=& -v_{F} \nabla\cdot\Psi^{\dagger}_{\bf p}(t){\bf \sigma}
\end{eqnarray}
since $\sigma^{\dagger}=\sigma$.
Multiplying (\ref{Dir}) with $\Psi^{\dagger}_{{\bf p}}(t)$
and (\ref{ConDir}) with $\Psi_{\bf p}(t)$ and adding them we obtain 
\begin{equation}
\frac{\partial}{\partial t}\left ( \Psi^{\dagger}_{\bf p}(t)\Psi_{\bf p}(t)\right ) =-v_{F} \nabla\cdot \left (\Psi^{*}_{\bf p}(t)\sigma\Psi_{\bf p}(t)\right )
\end{equation}
or, the graphino continuity equation:
\begin{equation}\label{ConEqu}
\frac{\partial}{\partial t}\rho(t) + \nabla\cdot {\bf j}(t) =0
\end{equation}
where
\begin{eqnarray}\label{Den}
\rho(t)&=& \Psi^{\dagger}_{\bf p}(t)\Psi_{\bf p}(t)\nonumber\\
& =& u^{(s)\dagger}_{{\bf p}(t)}u^{(s)}_{{\bf p}(t)}
\end{eqnarray}
is the probability density, and
\begin{eqnarray}\label{CurOpe}
{\bf j}(t)&=&v_{F}\Psi^{\dagger}_{\bf p}(t)\sigma\Psi_{\bf p}(t)\nonumber\\
&=& v_{F}u^{(s)\dagger}_{{\bf p}(t)}{\bf \sigma} u^{(s)}_{{\bf p}(t)}
\end{eqnarray}
is the probability current density.
We assume that the lower band (s= -1) is fully occupied and therefore in an adiabatic laser field harmonic emission is due essentially to the coherently driven graphino current in the upper band (s=+1), arising from the initially occupied states near the degenerate Dirac points.  
For a laser incident perpendicular to the graphene plane and linearly polarized along the x-axis, we substitute    
the spinors (\ref{TimDepSpi}) and (\ref{Ide2}) in Eq. (\ref{CurOpe}) and readily evaluate the quantum expectation value of the graphino current  operator to get: 
\begin{eqnarray}\label{CurXax}
{J}_{x}^{(s)}({\bf p}, t)&=& -e{j}_{x}(t)\nonumber\\
 &=& -e s v_{F}\cos{\phi_{\bf p}(t)}\nonumber\\
&=&-e s v_{F}\frac{p_{x}-\frac{e}{c}A_{x}(\eta)}{\sqrt{( p_{x}-\frac{e}{c}A_{x}(\eta))^{2} +p_{y}^{2}}}
\end{eqnarray}\\
Similarly, for the laser polarization along the y-axis we obtain
\begin{eqnarray}\label{CurYax}
{J}^{(s)}_{y}({\bf p},t) &=& -e s  v_{F}\sin{\phi_{\bf p}(t)}\nonumber\\
&=&-e s v_{F}\frac{p_{y}-\frac{e}{c} A_{y}(\eta)}{\sqrt{p_{x}^{2}+(p_{y}-\frac{e}{c}A_{y}(\eta))^{2}}}
\end{eqnarray} and for the more general case of an elliptically polarized laser field 
(cf. Eq.(\ref{VecPot})) we get \cite{foot1}:
\begin{eqnarray}\label{CurEll}
{\bf J}^{(s)}({\bf p},t) &=& -e s v_{F}\left (\hat{\epsilon}_{x}\cos{\phi_{\bf p}(t)} + \hat{\epsilon}_{y}\sin{\phi_{\bf p}(t)}\right )\nonumber\\
&=&-e s v_{F} \frac{\hat{\epsilon}_{x}(p_{x}- \frac{e}{c}A_{x}(\eta) ) +\hat{\epsilon}_{y}(p_{y}-\frac{e}{c} A_{y}(\eta))}
{\sqrt{(p_{x}-\frac{e}{c}A_{x}(\eta))^{2} + (p_{y} - \frac{e}{c} A_{y}(\eta))^{2}}}\nonumber\\
\end{eqnarray} where $A_{x}(\eta) =  A_{0}(\eta)\cos{(\xi/2)}\cos{\eta}$, and $A_{y}(\eta)= A_{0}(\eta)\sin{(\xi/2)}\sin{\eta}$.
\begin{figure}
\begin{centering}
\includegraphics[scale=0.34]{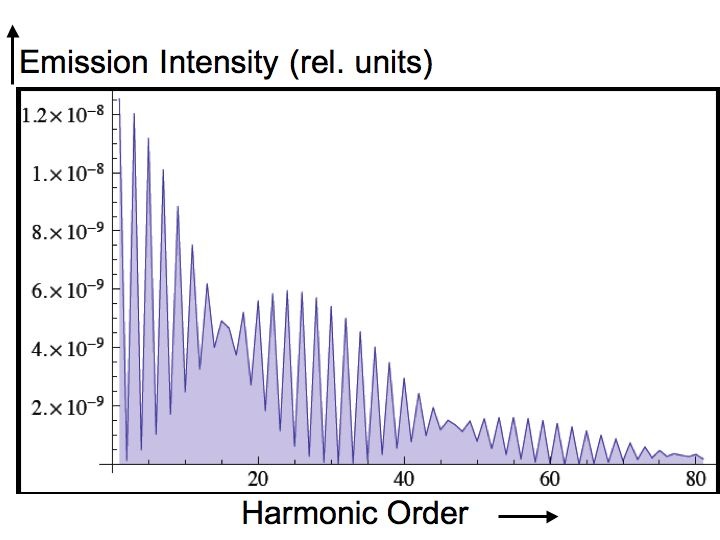}
\par\end{centering}
\caption{\label{Fig. 2} High order harmonic emission (shown up to the 13th harmonic) by THz laser interaction 
with ``graphinos'' near the Dirac point  $K(\frac{2\pi}{3a_{0}},\frac{2\pi}{3\sqrt{3}a_{0}})$;
illustrative calculation for $q_{x} a\equiv (K_{x}-k_{x})a=0.01$, $q_{y}b\equiv (K_{y}-k_{y})b=0.01$,
laser polarization along y-axis, frequency 3 THz, focused laser intensity I = 50 MW/cm$^{2}$. Note the presence of  the``revivals'' and ``plateaus''
in the middle and the end of distribution.}   
\end{figure} 
The intensity of the harmonic emission at the nth harmonic frequency $\Omega=n\omega$ 
depends essentially on the absolute square of the Fourier transform of the graphino current:
\begin{eqnarray}\label{Sig}
S^{(s)}(n\omega) &=& \left|(n\omega)\frac{1}{2\pi} \int_{-\pi}^{\pi}e^{-i n \eta}{\bf J}^{(s)}({\bf p},\eta) d\eta\right |^{2} 
\end{eqnarray}
where $n=1,2,3,\cdots$.
In Fig. 2 we present the results of calculations for the relative intensity of harmonics emitted from an occupied state in the neighborhood of the Dirac point K, by
a 3 THz laser with a focused intensity of 50 MW/cm$^{2}$, that is polarized linearly along the y-axis (cf. Fig. 1) and is directed perpendicular to the graphene plane. 
It should be noted that exactly at a Dirac point (e.g. $ K=(\frac{2\pi}{3a_{0}},\frac{2\pi}{3\sqrt{3}a_{0}}$)) the driven graphino current in the present case becomes a constant (cf. Eq. (\ref{CurYax})) and therefore no harmonics are produced by graphinos from this point. Nevertheless, as expected above high order harmonic emission can occur copiously from the initially occupied states in the upper band (s=+1) in the vicinity of the Dirac points. Fig. 2 shows an example of harmonic emission from a neighboring point ($q_{x}a=0.01, q_{y}b=0.01$), exhibiting highly non-perturbative relative intensity \cite{foot2} that is seen to decrease only about two orders of magnitude over some 81 orders of the harmonics. Moreover, it can seen clearly from the figure that the distribution of relative intensity is characterized by ``revivals''  and multiple ``plateaus'' within the range shown in Fig. 2  \cite{foot3}.

\subsection{Harmonic generation from graphene band}
The full dispersion relation of graphene is very different from that of the graphino quasi-particles. It should be therefore interesting to see the difference between the harmonic generation spectra in the two cases. We discuss below the results of our  
preliminary investigation of the difference in a particular case.
To this end we consider the energy band 
of  a graphene sheet in the tight-binding approximation (see, e.g. \cite{rev2}). In this approximation the upper and the lower bands are 
given 
\begin{eqnarray}
E^{(s)}(\bm{p}) &=& s t \{ 3 + 2 \cos{(2k_{y} b)} + 4\cos{(k_{x}a)}\cos{(k_{y}b)} \}^{\frac{1}{2}}\nonumber\\
&=& s t \{1+ 4\cos{(k_{x}a)}\cos{(k_{y}b)} + (2 \cos{(k_{y}b}))^{2}\}^{\frac{1}{2}}\nonumber\\
\end{eqnarray} 
The semiclassical band current  can be obtained by differentiating the band energy with respect to the band momentum and applying the
minimal coupling prescription. Thus for the band current induced by an elliptically polarized electromagnetic field propagating perpendicular to the graphene plane we find,
\begin{eqnarray}
\bm{J}^{(s)}(\bm{p},t)&=& \hat{\epsilon}_{x} j_{x}^{(s)}(\bm{k}(\eta)) +\hat{\epsilon}_{y} j_{y}^{(s)}(\bm{k}(\eta))\nonumber\\
 j_{x}^{(s)}(\bm{p},t) &= &s t (-2 a)\sin{(k_{x}(\eta)a)} \cos{(k_{y}(\eta)b)} \nonumber\\
 && \times \{1+ 4\cos{(k_{x}(\eta)a)} \cos{(k_{y}(\eta)b)} \nonumber\\
 && + (2 \cos{(k_{y}(\eta)b)})^{2} \}^{-\frac{1}{2}}\\
 j_{y}^{(s)}(\bm{p},t) &=& (-2 b)\{ \cos{(k_{x}(\eta)a)} \sin{(k_{y}(\eta)b)} \nonumber\\
 && + \cos{2(k_{y}(\eta)b)} \} \nonumber\\
&&\times \{1+ 4\cos{(k_{x}(\eta)a)}\cos{(k_{y}(\eta)b)} \nonumber\\
&& + (2 \cos{(k_{y}(\eta)b)})^{2})\}^{-\frac{1}{2}}\nonumber\\
\end{eqnarray}
where,
\begin{eqnarray}
\bm{k}(\eta) &=& (k_{x}(\eta),k_{y}(\eta))\nonumber\\
k_{x}(\eta) &=& k_{x} -\frac{e}{\hbar c}A_{x}(\xi)\cos{(\eta)}\\
k_{y}(\eta)&=& k_{y}-\frac{e}{\hbar c}A_{y}(\xi)\sin{(\eta)}
\end{eqnarray}
\begin{figure}
\begin{centering}
\includegraphics[scale=0.34]{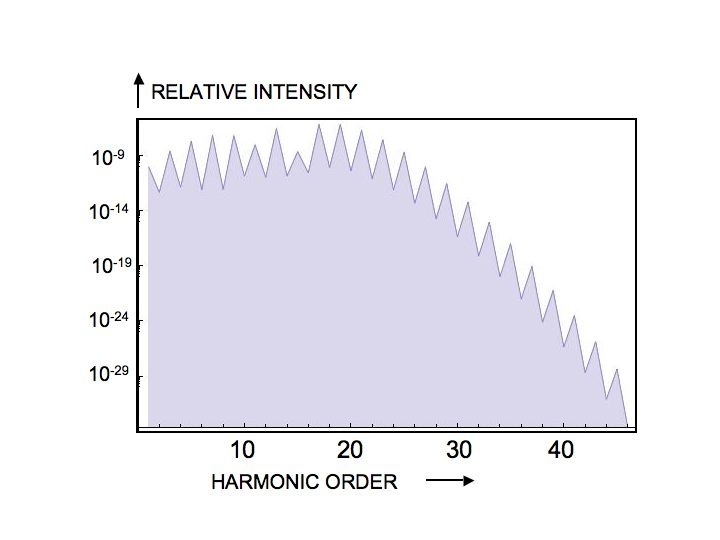}
\par\end{centering}
\caption{\label{Fig. 3} High order harmonic emission (shown up to the 46th harmonic) by aTHz laser interaction 
with a 2D graphene layer near the Dirac point  $K(\frac{2\pi}{3a_{0}},\frac{2\pi}{3\sqrt{3}a_{0}})$;
illustrative calculation for $q_{x}a \equiv (K_{x}-k_{x})a=0$, $q_{y}b\equiv (K_{y}-k_{y})b=0.01$,
laser polarization along y-axis, frequency 3 THz, focused laser intensity I = 500 GW/cm$^{2}$. Note the presence of  a cutoff 
order $n_{cutoff} \approx 20$ after which the relative harmonic intensity drops exponentially, consistent with the formula $n_{cutoff}\approx \frac{\omega_{0}}{\omega}=19.9$. 
}   
\end{figure} 
Let the laser be polarized along the y-axis and assuming that the initial states along $k_{y}$ and $k_{x}=0$ contribute the most to the driven current in this case we get:
 \begin{eqnarray}
j^{(s)}_{y}(p_{y},t) &=&  (-2 s t  b) \sin{(k_{y}(\eta)b)} \nonumber\\
&=& (-2 s t b) \sin{(k_{y} - \rho\sin{\eta})}
\end{eqnarray}
where,
$\rho=\frac{eF_{0}b}{\hbar \omega}$ and 
$F_{0}=\frac{ \omega A_{0}}{c}$.
 Using the Jacoby-Anger relation
 \begin{equation}
 e^{i \rho \sin{\eta}} = \sum_{n=-\infty}^{\infty} J_{n}(\rho) e^{i n\eta}
\end{equation}   
the Fourier transform of the current can be readily obtained:
 \begin{eqnarray}
 F.T. [J^{(s)}_{x}(p_{y},t)](n\omega) &=&(-2 s t b)\frac{1}{2 i}(e^{i k_{y} b} - (-1)^{n} e^{-ik_{y}b})\nonumber\\
&&\times (-1)^{n} J_{n}(\rho) \\ 
&=& (-2 s t b) \sin{( k_{y}b)} J_{n}(\rho)\nonumber\\
&& \mbox{for, n=even}\\
&=& (-2 s t b) i \cos{(k_{y}b)}J_{n}(\rho)\nonumber\\
&& \mbox{for, n=odd}
\end{eqnarray}
Using the above expression for the current in Eq. (\ref{Sig}), we obtain
for the relative intensity of the harmonics emitted from an initial state$ k_{y}$ in the s th. band analytically in the form:
\begin{eqnarray}
S_{x}^{(s)}(n\omega) &=& 4 t^{2} b^{2}(n\omega)^{2} J_{n}^{2}(\rho) \sin^{2}{(k_{y} b)},\nonumber\\
&& \mbox{for, n=even} \\
&=& 4 t^{2} b^{2}(n\omega)^{2} J_{n}^{2}(\rho) \cos^{2}{(k_{y} b)} \nonumber\\ 
&& \mbox{for, n=odd}
\end{eqnarray}
It is immediately seen from the formulae above that for $q_{y}=0$ only odd harmonics are generated while for non-zero valuesof $q_{y}$ both odd and even harmonics are emitted.    
From the known exponentially decreasing asymptotic behavior of Bessel functions of order n, for $n>\rho$, one also expects
an exponential ``cutoff'' of the harmonic emission above 
\begin{equation}
n_{cutoff} \approx \rho = \frac{\omega_{B}}{\omega}
\end{equation}
where $\omega_{B}\equiv \frac{eF_{0}b}{\hbar}$ is the so-called Bloch frequency.
Finally, in Fig. 3 we show the relative intensity distribution of the harmonics for the case of a 3 THz laser, polarized along the y-axis, at an intensity I = 500 GW/cm$^{2}$, near the Dirac point $(q_{x} a=0, q_{y} b=0.01)$.
As can be seen, now the harmonic spectrum extends significantly only up to about $n_{cutoff}\approx 20$ that confirms the prediction of the cutoff- formula  
$n_{cutoff} \approx \frac{\omega_{0}}{\omega}=19.9$; beyond this value it indeed falls of exponentially. This behavior is qualitatively different from that of the free graphinos discussed above (cf. Fig. 2), as well as from the case of free atoms and molecules for which a cutoff order exists but is given differently by the formula: $n_{cutoff}\approx I_{p} + 3.17 U_{p}$, where $U_{P}=\frac{e^{2}F_{0}^{2}}{4m\omega^{2}}$ is the so-called ponderomotive potential \cite{kulander,lewenstein}. The difference in the two cases of free graphinos and active electrons in a graphene layer is mainly due to the large excursion-radius in an intense field of the active particles which explore very different dispersion relations and hence different induced currents.   
In the case considered above, the relative intensity spectrum for 2D graphene is reminiscent of the high harmonic emission spectrum observed recently \cite{dimauro} in experiments with bulk ZnO crystals and analyzed within a rather similar 1D tight-binding model. 

\section{Conclusion}
We obtain a class of exact adiabatic solutions of the time-dependent Dirac equation for the low energy massless excitations of 2D graphene sheets or ``graphinos'' interacting with an electromagnetic field. 
They are expected to be useful for analyses of non-perturbative processes in intense laser fields.
As a first case of their use we predict copious emission of harmonic radiation by THz laser with considerable relative intensities.
Thus, for example, calculation near the Dirac point (e.g. $q_{x} a = q_{y} b = 0.01$) shows that many higher order harmonics can be generated by a linearly polarized  3 THz laser with a focused intensity of 50 MW/cm$^{2}$, whose relative intensities may change by only about two orders of magnitude from the 3rd to the 81st harmonic order. Furthermore the calculated distribution clearly shows two interesting domains of ``revival'' and ``plateau'' formation in the relative harmonic intensity. Preliminary results for a laser linearly polarized along the y-axis, and for initial states near the Dirac point show that high harmonic generation from a 2D graphene layer treated in a tight-binding approximation can be qualitatively different from the spectrum for the free graphinos. The former, instead, is rather analogous to the recently observed high harmonic spectrum from bulk ZO crystals.


\begin{thebibliography}{99}
\bibitem{foot-1}
For the sake of brevity as well as for the purpose of distinguishing the charged quasi-particles associated with the massless low energy excitations from the case of the 2D graphene sheet, and in analogy with the massless ``neutrinos'' we found it useful to refer to this new class of low energy zero-mass relativistic fermions (excited in graphene sheets) simply as ``graphinos''.
\bibitem{foot1}
It is interesting to note that the adiabatic Dirac current (\ref{CurEll}) is analogous to the non-relativistic semi-classical Schr\"odinger current in a uniform electric field (cf. \cite{mermin}).
\bibitem{foot2}
Probability of an n-photon perturbative process \cite{fai-book} decreases as 
(I/I$_{a}$)$^{n}$ with increasing n, where I is the laser intensity in W/cm$^{2}$ and I$_{a} (a.u.) \approx $3.51 W/cm$^{2}$.
\bibitem{foot3}
Details of high order harmonic emission from ``graphinos'' and 2D graphene sheets, and the wavelength, intensity and polarization dependence are to be reported elsewhere.
\end{thebibliography}
\end{document}